\documentclass{aastex631}
\graphicspath{{.}}
\usepackage{graphicx}
\usepackage{multirow}
\usepackage{amsmath}
\usepackage{float}
\DeclareUnicodeCharacter{2212}{-} 
\usepackage{threeparttable}
\begin{document}

\title{Hemispheric Asymmetry of Solar Active Regions Arises from a Nested Population}

\author[0000-0003-2622-7310]{A.A. Norton}
\affiliation{Physics Department, Stanford University, Stanford, CA 94305-4085}

\begin{abstract}
We investigate the longitude--time distribution of NOAA active regions (ARs) during Solar Cycles 22--24 and find statistically significant North--South asymmetry in AR emergence. Using an activity-nest identification algorithm, we show that this asymmetry is concentrated in the subset of ARs that participate in nests. Nest-member ARs exhibit substantially larger hemispheric asymmetry than either the full AR population or the non-nest population, and the asymmetry is largely removed when nest-member ARs are excluded. Monte Carlo tests with randomized longitudes and temporal perturbations show that the observed nesting and asymmetry exceed random expectations, implying that $\sim$6--18\% of ARs participate in a non-random, hemispherically asymmetric nesting component. This asymmetry is associated with temporally offset bursts of activity and distinct longitudinal clustering between the hemispheres, leading to reduced cross-equatorial coherence in the longitude--time distribution of solar ARs. Intervals of enhanced nesting activity and hemispheric asymmetry broadly coincide with enhanced hemispheric quasi-biennial variability and temporal evolution of the large-scale solar magnetic field, suggesting a possible connection between intermediate-timescale dynamo variability and the hemispheric organization of solar activity.
\end{abstract}

\keywords{Sunspots (1653) --- Solar cycle (1487) --- Solar dynamo (2001)}

\section{Introduction} 
Solar active regions (ARs) do not emerge randomly in longitude, but instead cluster within localized longitude–latitude bands over multiple rotations. These concentrations of activity, commonly referred to as activity nests, have been identified in sunspot catalogs, magnetograms, and coronal emission observations \citep[e.g.,][]{bumba1969,gaizauskas1983,Castenmilleretal1986,Bai1987,Henney2002,Gyenge2014,mandal:2017}. Such clustering implies that solar magnetic activity contains a non-axisymmetric component and may therefore provide insight into the operation of the solar dynamo.

Longitudinal concentrations of activity have also been described in terms of active longitudes, namely preferred longitudes at which magnetic activity repeatedly emerges over many rotations and, in some studies, over multiple solar cycles \citep{Bogart1982,Bai1987,Berdyugina2003,Usoskin2005}.Previous studies have further reported ``flip-flop'' behavior in which activity alternates between longitudinal bands separated by approximately $180^{\circ}$ \citep{Berdyugina2003,Usoskin2005}. However, the persistence and statistical significance of active longitudes remain debated, with some studies demonstrating that apparent longitudinal preferences can arise from analysis methods applied to stochastic emergence patterns \citep{pelt:2005}. In the present work, we do not search for persistent antipodal active longitudes. Instead, we investigate whether temporally intermittent activity nests give rise to hemispheric asymmetry in solar magnetic activity.

The importance of activity nests extends beyond the emergence properties of individual ARs. \citet{finley2024} showed that nested active regions can anchor the heliospheric current sheet— the warped extension of the solar magnetic equator into the heliosphere that separates opposite interplanetary magnetic polarities \citep{smith2001} —highlighting their potential influence on the evolution of the Sun's large-scale magnetic field. Other recent studies employing machine-learning and density-based clustering techniques similarly found that a substantial fraction of sunspot groups emerge within spatial–temporal nests \citep{Isik2023,karapinar2026}, and persistent hemispheric and longitudinal organization of flare-productive active regions has also been reported \citep{korsos2025}. Consistent with these findings, in a companion study (Paper I; \citet{Norton2025}) we found that approximately 40–50\% of AR magnetic flux during Solar Cycle 24 participates in nests that persist for several Carrington rotations, demonstrating that nesting is a robust and dynamically significant feature of solar magnetic activity.

Despite this progress, the hemispheric organization of nested activity remains poorly understood. In particular, it is unclear whether activity nests emerge coherently across the equator or occur preferentially within individual hemispheres. Addressing this question provides a new observational probe of hemispheric coupling and the degree of non-axisymmetry in the solar dynamo. In this Letter, we examine the longitudinal and temporal distribution of AR emergence during Solar Cycles 22–24. We quantify the statistical significance of hemispheric asymmetry in time–longitude space and determine the extent to which that asymmetry is concentrated within the subset of ARs participating in activity nests, as opposed to the remaining non-nested population.

Hemispheric organization may ultimately be interpreted within the framework of the Sun's large-scale parity structure. The global magnetic field can be described in terms of low-order spherical-harmonic families whose relative amplitudes evolve over the solar cycle \citep{Bullard1954,Moffatt1978,DeRosa:2012}. Because mixed-parity dynamo states can produce hemispheric dominance \citep{Gallet2009}, the hemispheric organization of activity nests may provide an observational constraint on the evolving balance between large-scale magnetic-field families.

\begin{figure}[h]
    \centering  
    \includegraphics[trim=18 33 0 25, clip, width=0.735\textwidth]{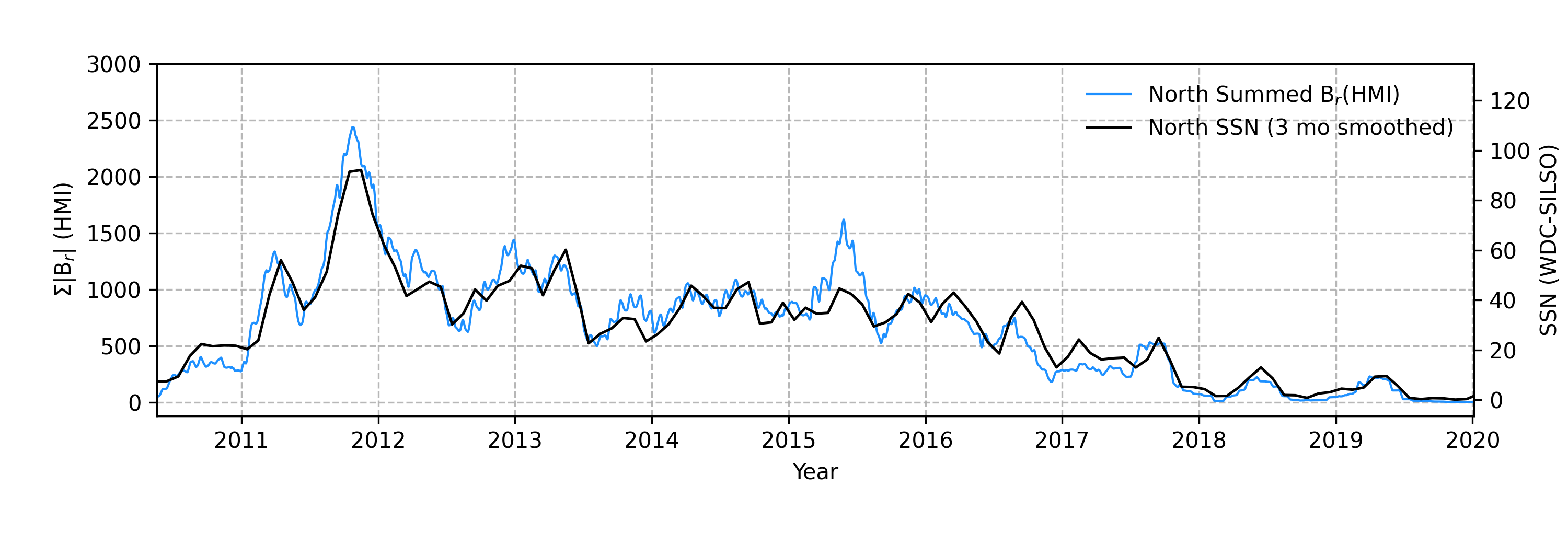}
    \includegraphics[trim=-1 13 0 5, clip, width=0.745\textwidth]{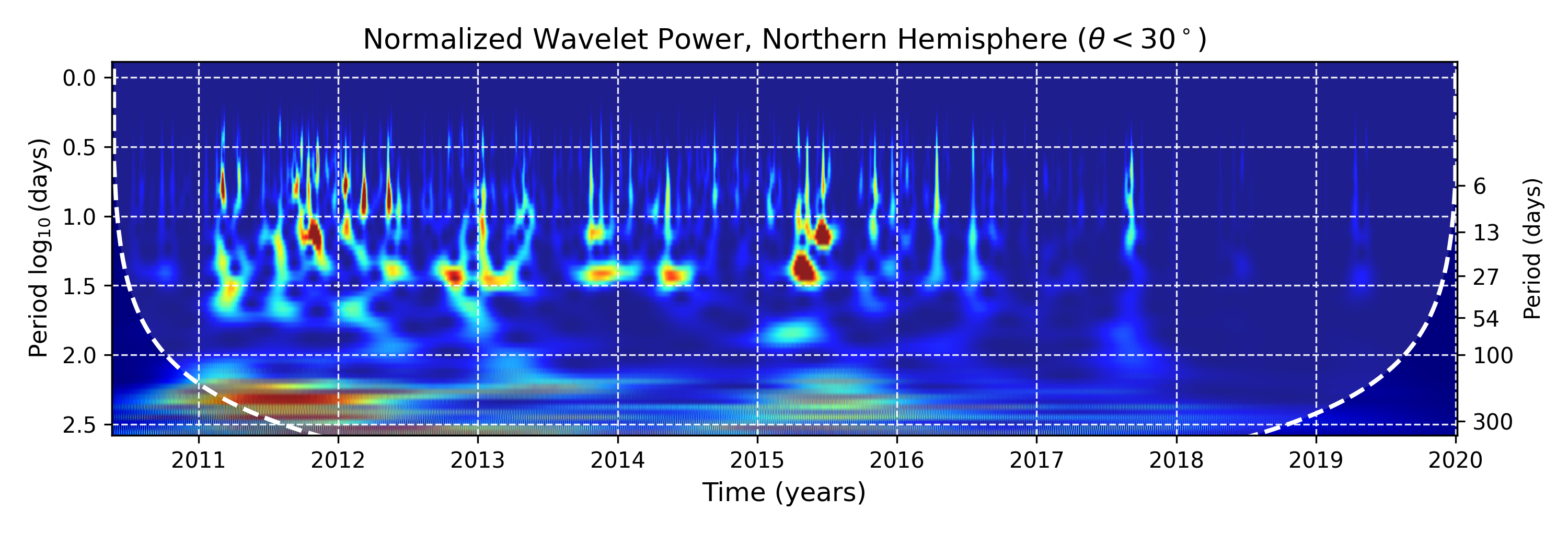} 
    \includegraphics[trim=-1 13 0 5, clip, width=0.745\textwidth]{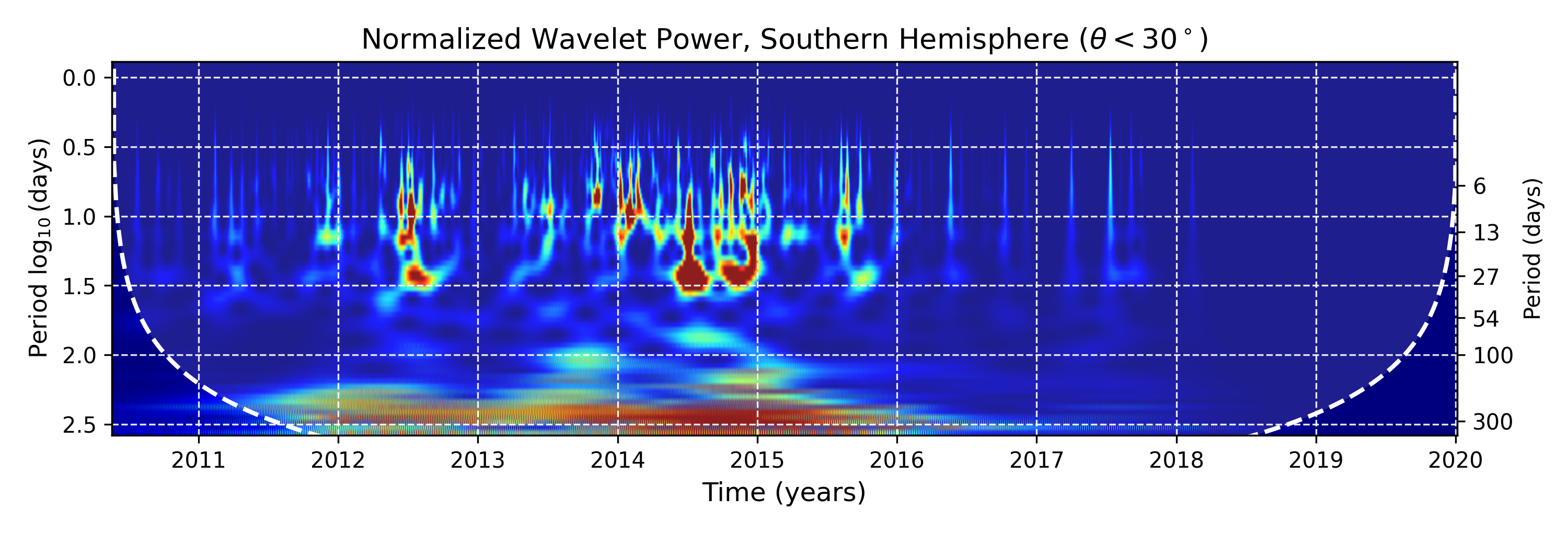}
    \includegraphics[trim=18 33 0 25, clip, width=0.735\textwidth]{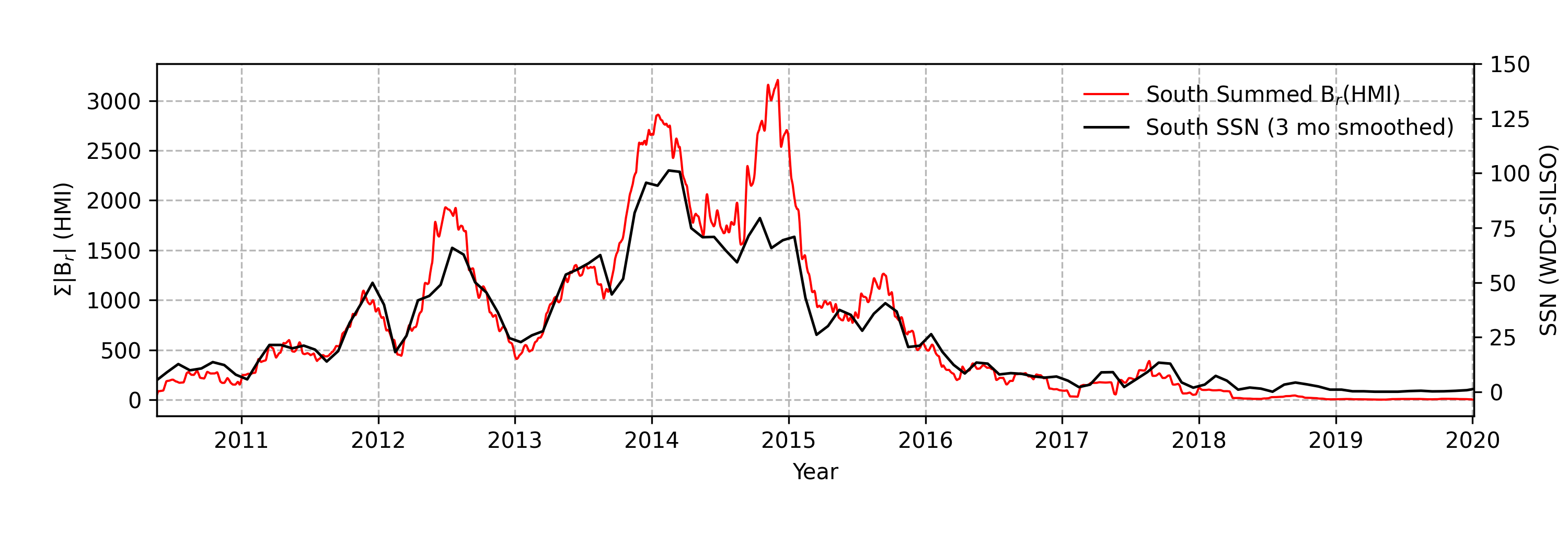}
    \includegraphics[trim=18 33 0 25, clip, width=0.735\textwidth]{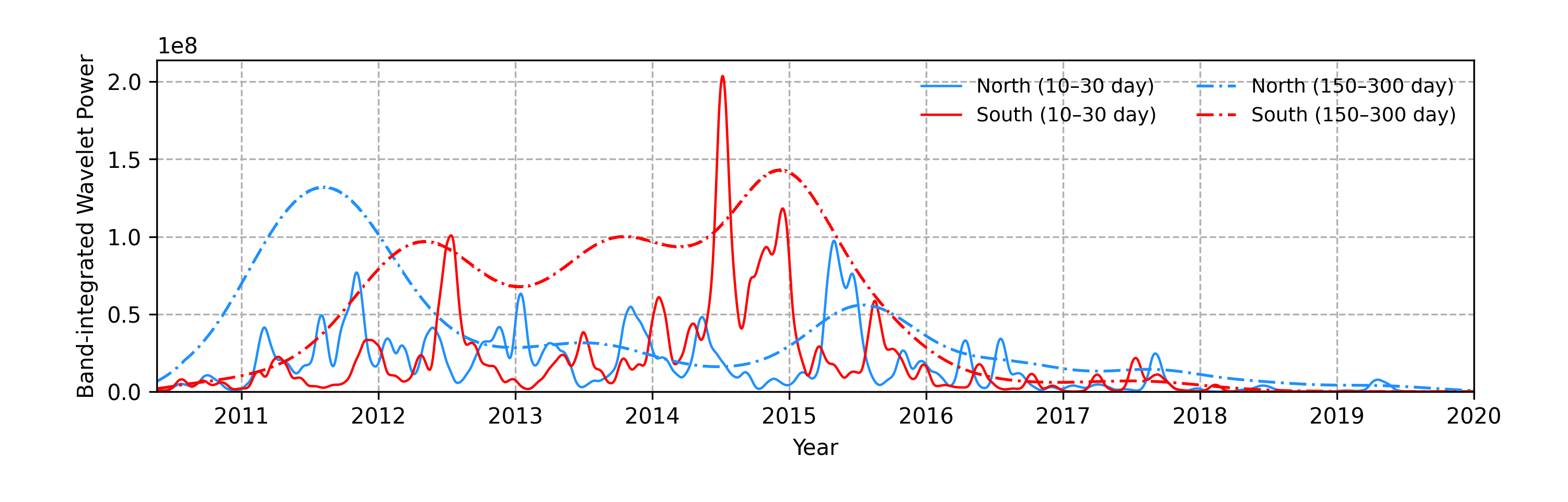}
    \caption{Monthly hemispheric sunspot number (SSN) and summed HMI $B_r$, boxcar smoothed over 3 months, for the North (top panel) and South (fourth panel) hemispheres. The second and third panels show the corresponding Morlet wavelet power spectra for periods between 1--379 days in the North and South hemispheres, respectively. Dark blue (red) indicates low (high) power. Enhanced wavelet power occurs in temporally localized emergence episodes that are largely offset between the two hemispheres. The fifth panel isolates this power in two period ranges -- 10--30 days (solid) and 150--300 days (dash-dot) -- for the North (blue) and South (red). In the 150--300 day range, the North shows a single burst of power in 2011, while the South exhibits persistently enhanced power from 2012 through 2015. In the 10--30 day range, both hemispheres show a series of discrete bursts that alternate in time between hemispheres.}
    \label{fig:fig1}
\end{figure}

\section{Data and Methods}

We analyze NOAA and Royal Greenwich Observatory (RGO) active-region (AR) data for Solar Cycles 22--24 using AR positions and areas compiled and distributed by NOAA’s National Centers for Environmental Information (NCEI) \citep{RGO_sunspots, NOAA_NCEI_sunspots}. Each AR is sampled only once, at the time closest to central meridian passage, to avoid duplicate counting and minimize projection effects. Hemispheric asymmetry is examined using the longitude--time distribution of AR emergence.

To characterize the temporal evolution of hemispheric activity, we use synoptic Carrington-rotation maps of the radial magnetic field, $B_r$, from the Helioseismic and Magnetic Imager (HMI) aboard the Solar Dynamics Observatory \citep{Scherrer2012, Schou2012}. We compute hemispheric time series by summing the absolute value of $B_r$ between $0^{\circ}$--$35^{\circ}$ latitude in each hemisphere. Time--frequency evolution is analyzed using the continuous Morlet wavelet transform \citep{torrence1998practical}, allowing identification of episodic hemispheric bursts of intermediate-timescale activity.

Activity nests are identified using the automated procedure developed in Paper I \citep{Norton2025}, based on the criteria of \citet{Castenmilleretal1986}, in which at least three ARs occur within four Carrington rotations and within $\pm7.5^{\circ}$ longitude and $\pm5^{\circ}$ latitude of one another. We search for nests across a range of rotation rates guided by the long-lived group (LLG) rotation law of \citet{Nagovitsyn2018, Nagovitsyn2023},
\begin{equation}
\omega = 14.3499 - 2.86\sin^2\lambda,
\end{equation}
and additionally search for retrograde and prograde nesting rates between 405--435 nHz (synodic). Nest-member ARs are compared with the full AR population and with the remaining non-nest population.

To quantify hemispheric asymmetry, AR locations are dilated by $\pm10^{\circ}$ in longitude and $\pm1$ Carrington rotation in time. ARs whose dilated regions intersect with ARs in the opposite hemisphere are classified as symmetric, while those without counterparts are classified as asymmetric. Statistical significance is evaluated using Monte Carlo realizations in which AR longitudes are randomized and emergence times are shifted by up to $\pm2$ Carrington rotations while preserving the overall activity level.

To examine the large-scale magnetic-field structure, we analyze the temporal evolution of the signed axial ($m=0$) harmonic coefficients for the dipole ($\ell=1$), quadrupole ($\ell=2$), and octupole ($\ell=3$) components derived from HMI synoptic maps for Solar Cycle 24 and from Wilcox Solar Observatory (WSO) synoptic maps for Solar Cycles 22 and 23 \citep{hoeksema:2014}.

Monthly hemispheric sunspot area was calculated separately for the northern and southern hemispheres using the NOAA/Greenwich AR catalog, counting each AR only once at the time closest to central meridian crossing. To isolate intermediate-timescale variability associated with QBO-like behavior, a 24-month running mean was subtracted from the monthly hemispheric sunspot-area time series. The resulting residuals were smoothed with a 5-month running mean and used as hemispheric QBO proxies.

\begin{figure}[H]
    \centering  
    \includegraphics[trim=8 20 0 0, clip, width=0.85\textwidth]{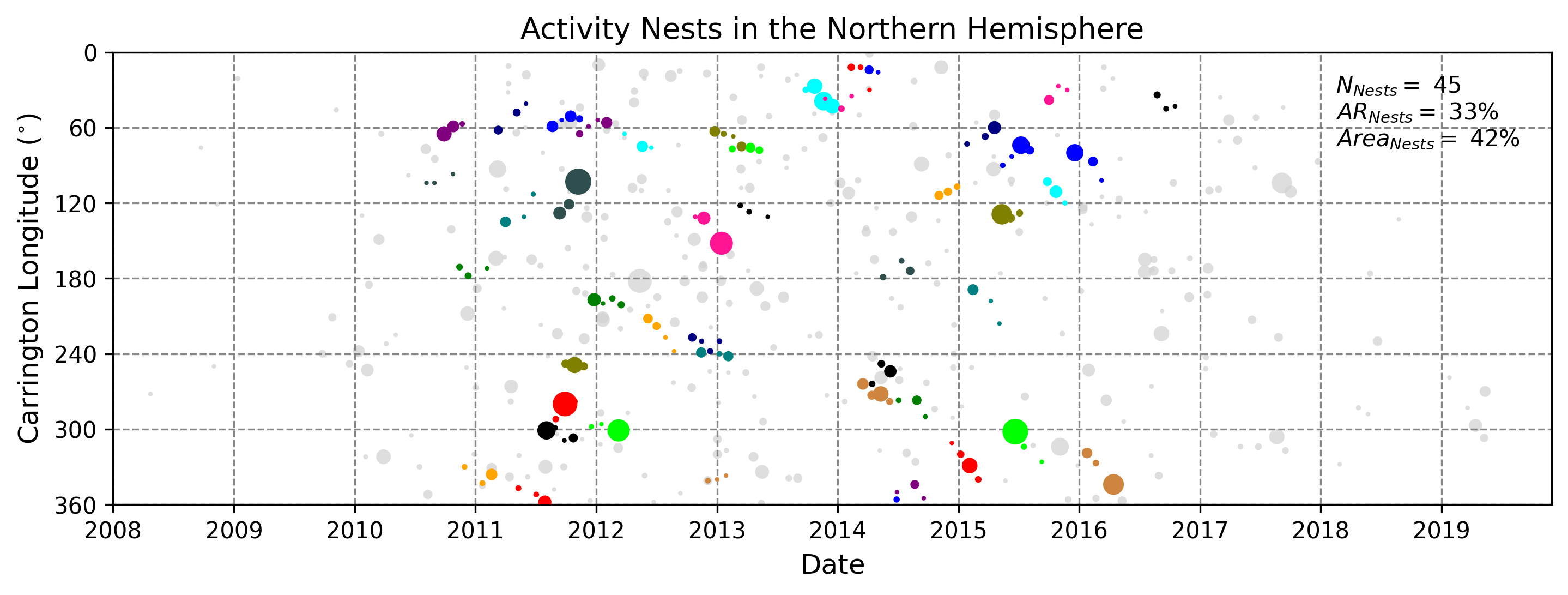}
     \includegraphics[trim=8 20 0 0, clip, width=0.85\textwidth]{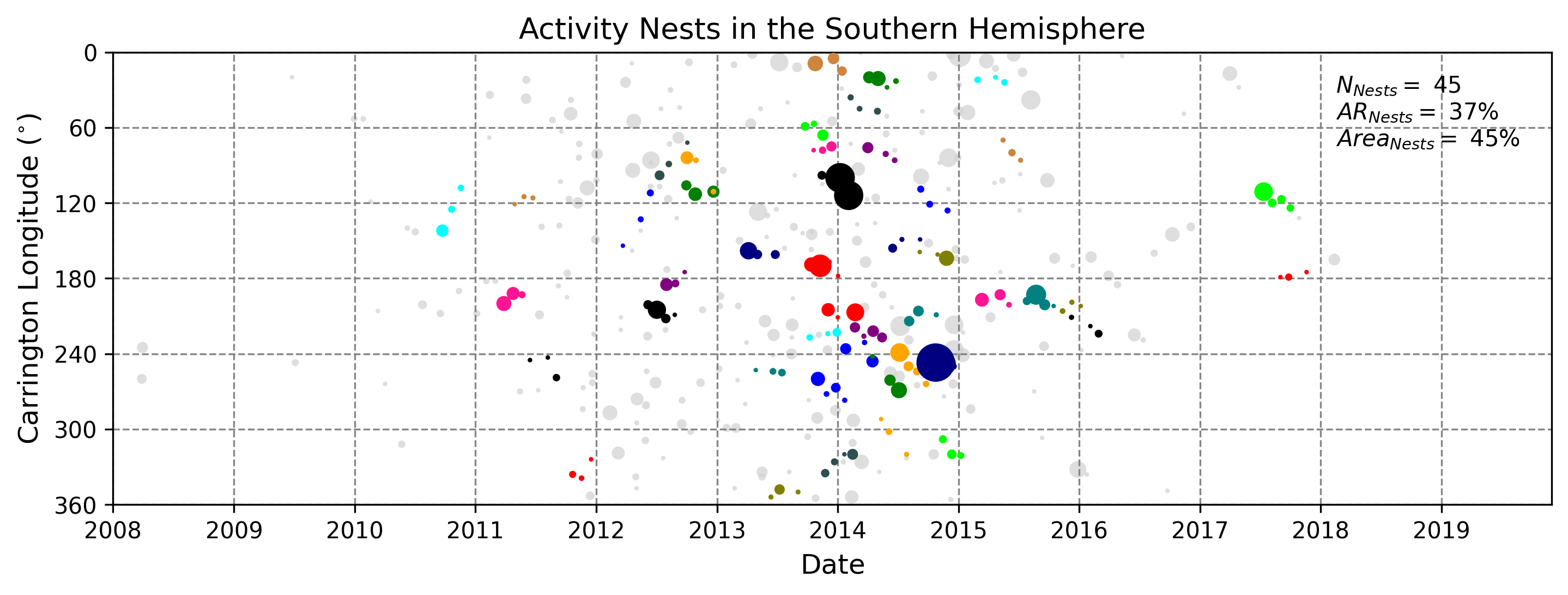} 
    \caption{Nest-member ARs during Solar Cycle 24 shown in longitude--time coordinates for the North (top) and South (bottom) hemispheres. ARs are sampled nearest central meridian passage and restricted to areas greater than 50~$\mu$Hem. Symbol size is proportional to AR area, while ARs belonging to the same nest are plotted with the same color. ARs that do not participate in nests are shown as light gray symbols in the background. The hemispheres exhibit distinct bursts of nested activity that occur at different times and longitudes. The corresponding figures for Solar Cycles 22 and 23 are available as an online Figure Set.}
    \label{fig:fig2}
\end{figure}
\figsetstart
\figsetnum{2}
\figsettitle{Nest-Member ARs in Longitude--Time Coordinates}
\figsetgrpstart
\figsetgrpnum{2.1}
\figsetgrptitle{Solar Cycle 22}
\figsetplot{FigSet2_N22.png}
\figsetplot{FigSet2_S22.png}
\figsetgrpnote{Nest-member ARs during Solar Cycle 22 shown in longitude--time coordinates for the North and South hemispheres, in the same format as Figure 2.}
\figsetgrpend
\figsetgrpstart
\figsetgrpnum{2.2}
\figsetgrptitle{Solar Cycle 23}
\figsetplot{FigSet2_N23.png}
\figsetplot{FigSet2_S23.png}
\figsetgrpnote{Nest-member ARs during Solar Cycle 23 shown in longitude--time coordinates for the North and South hemispheres, in the same format as Figure 2.}
\figsetgrpend
\figsetend

\begin{figure}[t]
    \centering  
    \includegraphics[trim=8 20 0 0, clip, width=0.94\textwidth]{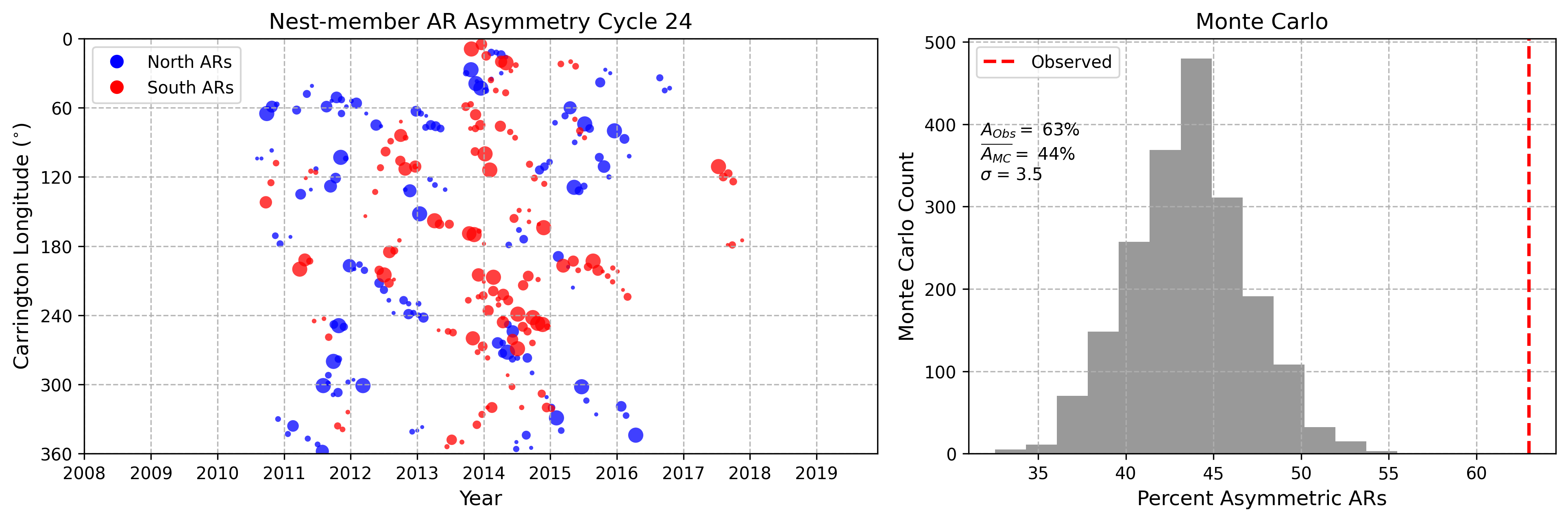}
    \includegraphics[trim=8 20 0 0, clip, width=0.94\textwidth]{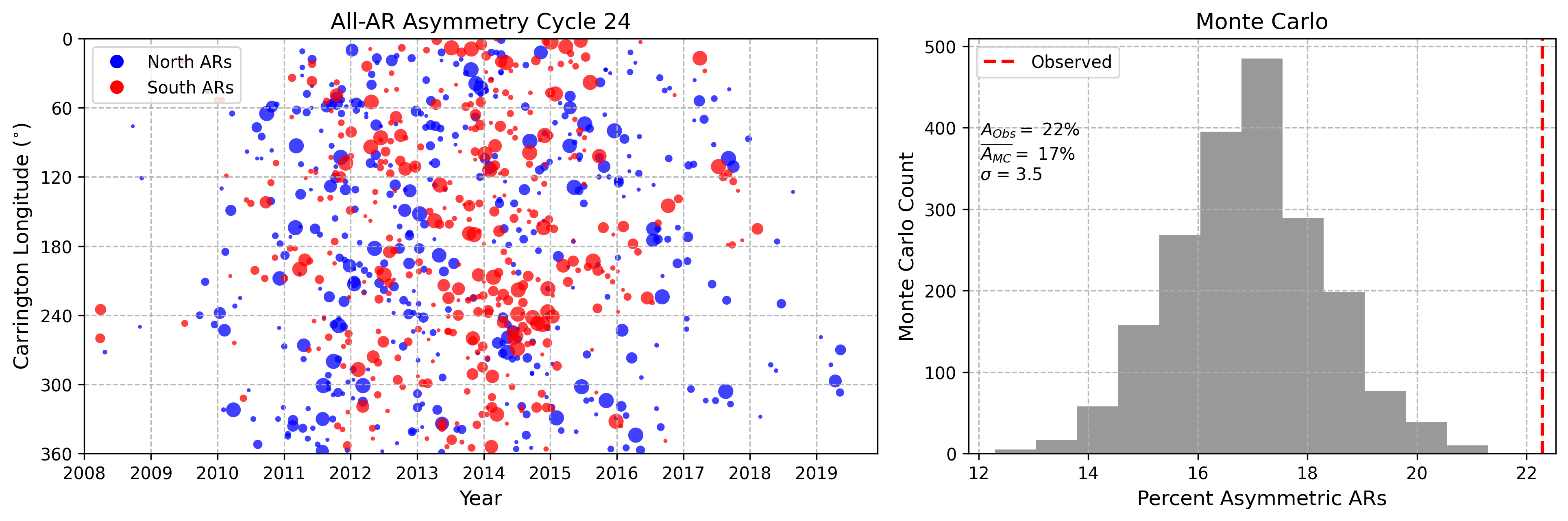}
    \includegraphics[trim=8 5 0 0, clip, width=0.94\textwidth]{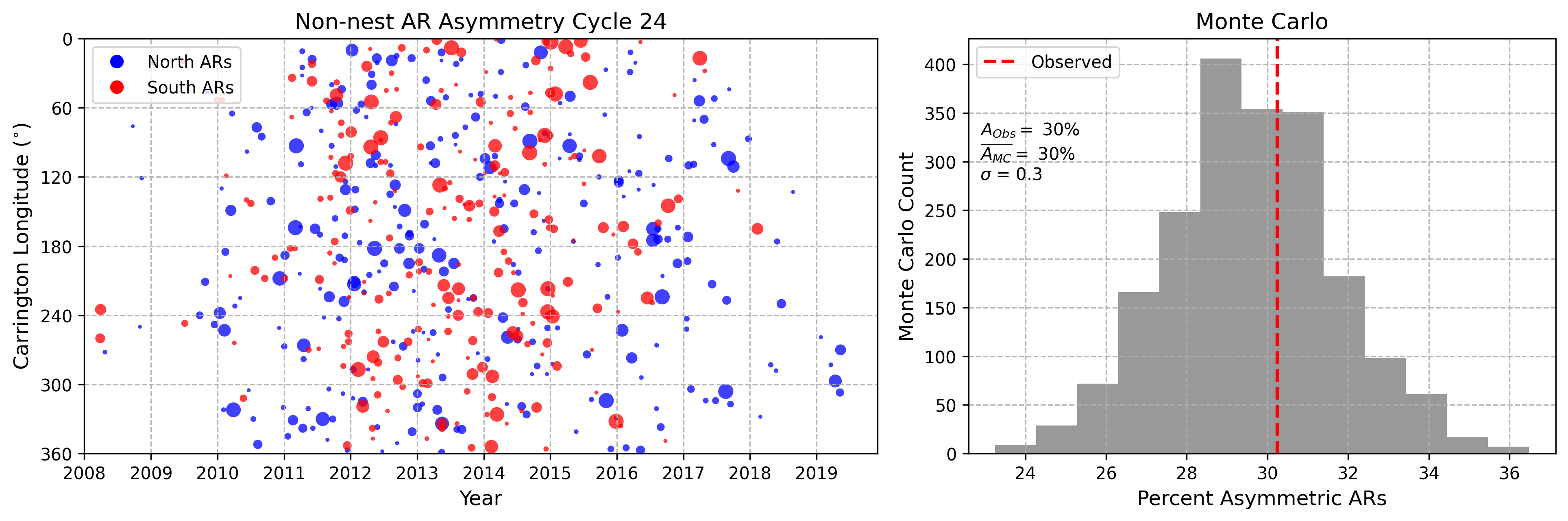}
    \caption{North--South hemispheric asymmetry of Solar Cycle 24 AR emergence in longitude--time coordinates (left) and histogram of asymmetry in the Monte Carlo realizations (right) compared to observed asymmetry. North (South) ARs are shown in blue (red) for the nest-member ARs (top), full AR population (middle), and the remaining non-nest population (bottom panels). AR locations are dilated by $\pm10^{\circ}$ in longitude and $\pm1$ Carrington rotation in time when determining cross-equatorial overlap (dilation not shown). ARs whose dilated regions intersect with ARs in the opposite hemisphere are classified as symmetric, while those without counterparts are classified as asymmetric. The observed asymmetry ($A_{\rm Obs}$, dashed red line) is compared with Monte Carlo realizations using randomized longitudes and temporal shifts of up to $\pm2$ Carrington rotations (right panel). Significant asymmetry is present in the nest-member and all-AR populations ($3.5\sigma$), while the non-nest population is statistically consistent with random symmetry in the Monte Carlo realizations ($0.3\sigma$).}
    \label{fig:fig3}
\end{figure}

\section{Results}
\subsection{Activity Nests and Temporal Bursts}

The time series of HMI $B_r$, summed between $0^{\circ}$--$35^{\circ}$ latitude in each hemisphere and smoothed with a 3-month boxcar, are shown in the top and fourth panels of
Figure~\ref{fig:fig1} for the northern and southern hemispheres, respectively. Monthly hemispheric sunspot numbers from WDC-SILSO, smoothed over the same interval, are also shown. The close correspondence between summed $B_r$ and hemispheric sunspot number indicates that integrated magnetic flux provides a reliable measure of hemispheric activity. The second and third panels of Figure~\ref{fig:fig1} show the corresponding normalized Morlet wavelet power spectra for the northern and southern hemispheres, respectively. Intervals of enhanced intermediate-timescale power occur at different
times in the two hemispheres.

To distinguish the timescales over which hemispheric power differs, the fifth panel isolates the band-integrated wavelet power in two period ranges, 10--30~days and 150--300~days, for each hemisphere separately. In the 150--300~day (quasi-biennial) range, the northern hemisphere exhibits a single pronounced burst of power in 2011, whereas the southern hemisphere shows persistent, elevated power extending from $\sim$2012 through $\sim$2015, with a broad maximum near 2012.5 that remains elevated rather than declining before rising to a second, stronger maximum near 2015. In the 10--30~day range, both hemispheres exhibit a series of discrete bursts that alternate in time between hemispheres: the northern hemisphere shows enhanced power in 2011, in late 2012 and late 2013, and again in early 2015, culminating in its largest short-timescale burst in mid-2015; the southern hemisphere shows enhanced power in mid-2012, intermittently through 2013--2014, and a pronounced peak in mid-2014 that is the strongest short-timescale feature in either hemisphere, followed by an additional burst in early 2015. These alternating short-timescale bursts, together with the persistent offset in longer-timescale power, indicate that intermediate-timescale magnetic variability in the two hemispheres oscillates out of phase over substantial portions of the cycle, rather than occurring as isolated, unrelated episodes.
\begin{table*}
\centering
\begin{threeparttable}
\caption{AR Nesting, Asymmetry, and QBO-Nest Statistics for Cycles 22--24}
\label{tab:nest_asymmetry}
\begin{tabular}{lcccccc}
\hline\hline
 & \multicolumn{2}{c}{Cycle 24} & \multicolumn{2}{c}{Cycle 23} & \multicolumn{2}{c}{Cycle 22} \\
\cline{2-3} \cline{4-5} \cline{6-7}
Quantity & Obs & MC & Obs & MC & Obs & MC \\
\hline
$N_{\rm ARs, Total}$ & 821 &-  &1371 &-  & 1171 & - \\
$N_{\rm nests}$ & \textbf{90} & 49 & \textbf{178} & 108  & \textbf{144}  & 84  \\
%$N_{\rm nests, N/S}$ & 45/45 & - & 82/96 & - & 65/79 & - \\
ARs in Nests (\%) & \textbf{36} & 18 & \textbf{42} & 24 &  \textbf{40}&  22\\
%Excess ARs in Nests (\%)& \textbf{18}& - & \textbf{18} & - & \textbf{18} & - \\
AR Area in Nests (\%)& \textbf{44} & 20 & \textbf{55} & 25 & \textbf{51} &24  \\
%Excess ARs in Nests (\%)& \textbf{24}& - & \textbf{30} & - & \textbf{27} & - \\
\hline
\multicolumn{7}{c}{\textit{Asymmetry Analysis}} \\
%Asymmetry Analysis & & & & & &\\
Nested AR Asymmetry (\%) & \textbf{63} & 44 & \textbf{46} & 26 &  \textbf{37}&  23\\
All AR Asymmetry (\%) & \textbf{22} & 17 & \textbf{13}$^{\dagger}$ & 11$^{\dagger}$ & \textbf{13} & 9 \\
Non-nest AR Asymmetry (\%) & \textbf{30}$^{\S}$ & 30$^{\S}$  & \textbf{21}$^{\S}$ & 21$^{\S}$ & \textbf{24}$^{\ddagger}$&  21$^{\ddagger}$\\
\hline
\multicolumn{7}{c}{\textit{QBO--Number of Nests Correlation}} \\
North ($r$) & \multicolumn{2}{c}{0.54} & \multicolumn{2}{c}{0.29} & \multicolumn{2}{c}{0.46} \\
South ($r$) & \multicolumn{2}{c}{0.46} & \multicolumn{2}{c}{0.41} & \multicolumn{2}{c}{0.39} \\
\hline
\end{tabular}
\begin{tablenotes}
\footnotesize
\item MC values represent the mean of the Monte Carlo realizations. 
%\item ``Excess'' quantities are defined as the difference between the observed values and the mean Monte Carlo values.
\item Statistical significance relative to Monte Carlo realizations exceeds $3\sigma$ unless otherwise noted:
$^{\dagger}$ ($2 \le \sigma < 3$),
$^{\ddagger}$ ($1 \le \sigma < 2$),
and $^{\S}$ ($\sigma < 1$).
\item Pearson correlation coefficient ($r$) between hemispheric QBO residuals (detrended sunspot area) and activity-nest counts, computed separately for each hemisphere and cycle, are all statistically significant ($p < 10^{-3}$). 
\end{tablenotes}
\end{threeparttable}
\end{table*}

Activity nests identified in Solar Cycle 24 are shown in Figure 2 for the northern and  southern hemispheres. Nests were identified using the criteria of \citet{Castenmilleretal1986},  requiring at least three ARs within four Carrington rotations and within $\pm7.5^{\circ}$ longitude and $\pm5^{\circ}$ latitude of one another. Rotation rates between 405--435 nHz  were examined. Using these criteria, we identify 90 activity nests in Cycle 24 (45 per hemisphere), comprising 36\% of the AR population and 44\% of the total AR area (Table~1, rows 3 and 4 Obs entries). The nests occur in temporally localized bursts and occupy distinct longitude bands in the two hemispheres. Monte Carlo realizations with randomized longitudes and temporal perturbations produce substantially smaller nesting fractions, with only 18\% of ARs and 20\% of AR area participating in nests (Table~1, rows 3 and 4 MC entries). The observed number of ARs in nests in Cycle 24 therefore exceeds random expectations by approximately 18\%, indicating that a distinct fraction of AR emergence occurs in non-random clustered configurations.

A similar pattern is found in Solar Cycles 22 and 23 (Table~1); the corresponding longitude--time distributions of nest-member ARs for these two cycles are provided as an online Figure Set accompanying Figure~2. Cycle 23 contains 178 nests comprising 42\% of ARs and 55\% of AR area, while Cycle 22 contains 144 nests comprising 40\% of ARs and 51\% of AR area. In both cycles, Monte Carlo realizations yield substantially smaller nesting fractions, again implying that the observed nesting fraction is significant and suggesting that a persistent fraction of solar AR emergence is governed by non-random longitudinal and temporal clustering independent of cycle amplitude.

\subsection{Hemispheric Asymmetry}

The hemispheric asymmetry of AR emergence is illustrated in Figure~\ref{fig:fig3} for the nest-member population, the all-AR population, and the remaining non-nest population. The nest-member AR population exhibits the strongest and most statistically significant hemispheric asymmetry in all three cycles (Table~\ref{tab:nest_asymmetry}). In Cycle 24, 63\% of nest-member ARs are asymmetric, compared to 44\% in the Monte Carlo realizations, corresponding to a significance exceeding $3\sigma$. Similar statistically significant departures from the Monte Carlo realizations are present in Cycles 23 and 22, with observed asymmetries of 46\% and 37\%, respectively, with significance exceeding $3\sigma$. 

As an additional test of robustness, we verified that this result is not an artifact of the smaller sample size associated with nest-member ARs. Random subsamples of size $N_{\rm nest}$ (147 North, 145 South ARs) drawn from the non-nest population alone yield a mean asymmetry of 47\%, confirming that smaller samples inflate this metric relative to the full non-nest population (30\%; Table~\ref{tab:nest_asymmetry}). However, the observed nest-member asymmetry of 63\% remains $\sim 3.5\sigma$ above this size-matched null distribution, indicating that the elevated asymmetry of nest-member ARs cannot be explained by sample size alone.

\begin{figure}[H]
    \centering  
    \includegraphics[trim=0 0 0 0, clip, width=0.835\textwidth]{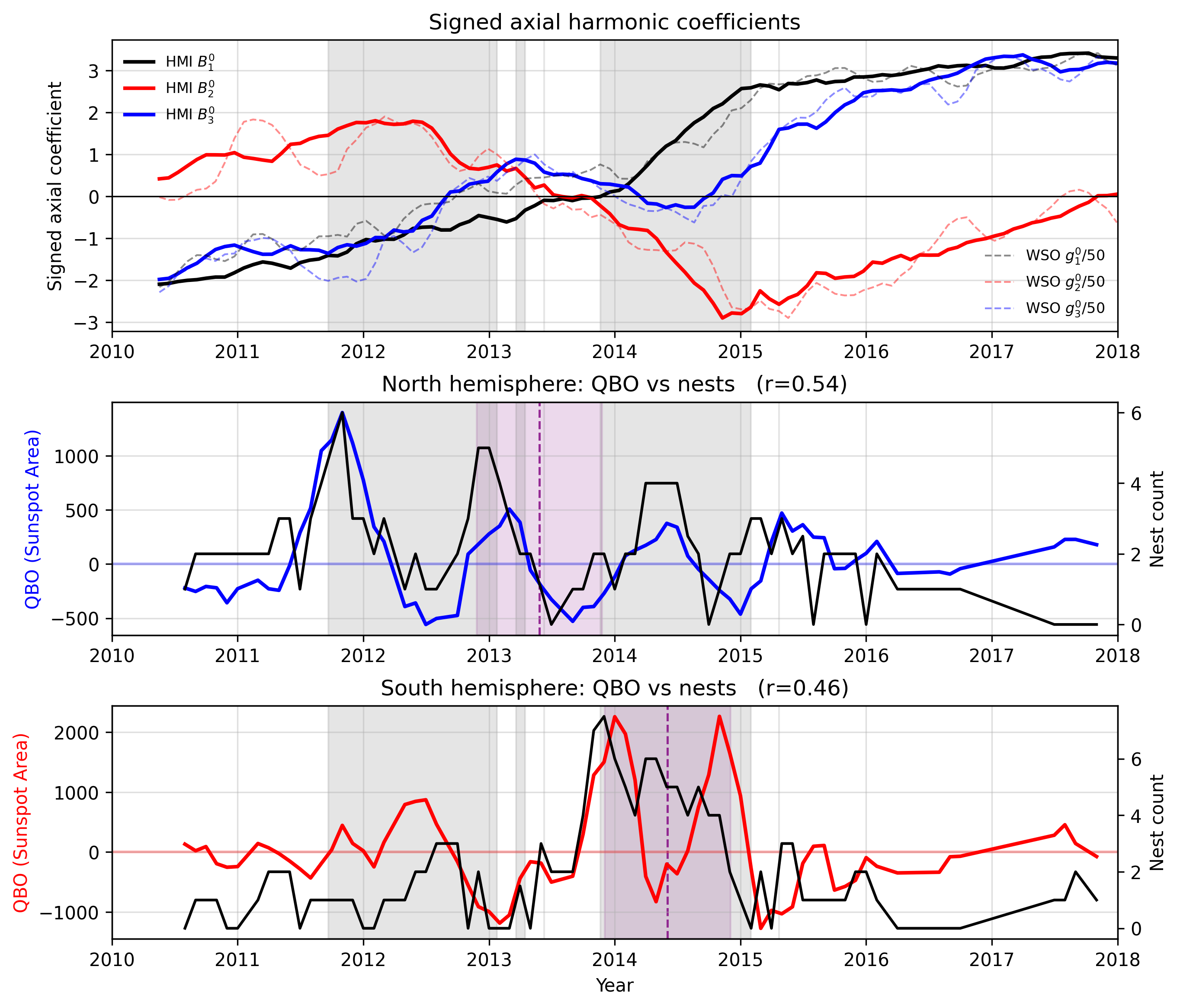}
    \caption{Top panel: Temporal evolution of the signed axial harmonic coefficients during Solar Cycle 24. Solid curves show the HMI coefficients $B_1^0$ (black), $B_2^0$ (red), and $B_3^0$ (blue), corresponding to the axial dipole, quadrupole, and octupole components, respectively, computed using orthonormal spherical harmonics following the convention of \citet{DeRosa:2012}. Dashed curves show the corresponding WSO coefficients $g_1^0$, $g_2^0$, and $g_3^0$, which use the Schmidt semi-normalization convention \citep{DeRosa:2012} and are rescaled (divided by 50) for visual comparison only; the WSO curves are shown as a qualitative, independent cross-check of the temporal behavior and are not used in any quantitative comparison. Gray shading indicates intervals during which the magnitude of the HMI axial quadrupole coefficient exceeds that of the HMI axial dipole coefficient, $|B_2^0| \geq |B_1^0|$, using the internally consistent HMI coefficients only. Purple shading and dashed lines indicate the approximate intervals and midpoints of the northern and southern polar-field reversals \citep{pishkalo2019}. Middle and bottom panels: Hemispheric QBO residuals derived from hemispheric sunspot area are shown for the North (blue; middle panel) and South (red; bottom panel) hemispheres together with the corresponding activity-nest counts (black). The northern and southern hemispheres exhibit temporally offset bursts of QBO activity and nesting. Intervals of enhanced nesting activity broadly overlap intervals in which the axial quadrupole coefficient becomes comparable to or exceeds the axial dipole coefficient. The corresponding figures for Solar Cycles 22 and 23 are available as an online Figure Set.}
    \label{fig:quadupole}
\end{figure}
\figsetstart
\figsetnum{4}
\figsettitle{Signed Axial Harmonic Coefficients and Hemispheric QBO--Nest Comparison}
\figsetgrpstart
\figsetgrpnum{4.1}
\figsetgrptitle{Solar Cycle 22}
\figsetplot{FigSet4_Cycle22.png}
\figsetgrpnote{Temporal evolution of the signed axial harmonic coefficients, hemispheric QBO residuals, and activity-nest counts for Solar Cycle 22, in the same format as Figure 4.}
\figsetgrpend
\figsetgrpstart
\figsetgrpnum{4.2}
\figsetgrptitle{Solar Cycle 23}
\figsetplot{FigSet4_Cycle23.png}
\figsetgrpnote{Temporal evolution of the signed axial harmonic coefficients, hemispheric QBO residuals, and activity-nest counts for Solar Cycle 23, in the same format as Figure 4.}
\figsetgrpend
\figsetend

Surprisingly, statistically significant time-longitude hemispheric asymmetry is also present in the full AR population (middle panel, Figure~\ref{fig:fig3}), indicating that the asymmetry associated with activity nests is sufficiently strong to remain detectable even when combined with the larger non-nest population. The asymmetry amplitudes are substantially smaller than those of the nest-member population, but remain significant relative to the Monte Carlo realizations. In Cycle 24, the observed asymmetry is 22\%, compared to 17\% in the Monte Carlo realizations, corresponding to a significance exceeding $3\sigma$. Cycles 23 and 22 exhibit similarly modest but statistically significant departures from the randomized populations (Table~\ref{tab:nest_asymmetry}).

After removing nest-member ARs, the remaining non-nest population becomes statistically consistent with the Monte Carlo realizations. In Cycle 24, the observed asymmetry of the non-nest population is 30\%, identical to the Monte Carlo expectation, with significance below $1\sigma$. Although the raw asymmetry percentage of the non-nest population is larger (30\%) than that of the full AR population (22\%; Table~\ref{tab:nest_asymmetry}), this does not indicate stronger hemispheric organization. Rather, removal of the nest-member ARs reduces the total number of ARs and increases the asymmetry expected from randomized realizations alone, resulting in no statistically significant asymmetry above the Monte Carlo expectation in the non-nest population.

These results demonstrate that the statistically significant hemispheric asymmetry of solar AR emergence is concentrated within the subset of ARs participating in nests. The disappearance of significant asymmetry after removing nest-member ARs indicates that the background AR emergence process is broadly consistent with random hemispheric occurrence, while a smaller subset of ARs emerges in temporally and longitudinally localized hemispheric bursts.

The difference between the observed and Monte Carlo fraction of ARs in nests is 18\% and provides a conservative upper bound on the fraction of ARs participating in a hemispherically asymmetric nested component, since not all nested ARs necessarily contribute directly to hemispheric asymmetry. A conservative lower bound can be estimated from the statistically significant asymmetry of the nest-member population above the Monte Carlo realizations,
\begin{equation}
f_{\rm nest}(A_{\rm nest,Obs}-A_{\rm nest,MC}),
\end{equation}
where $f_{\rm nest}$ is the observed fraction of ARs participating in nests. This yields approximately 7\%, 8\%, and 6\% of the total AR population for Cycles 24, 23, and 22, respectively. Together, these estimates suggest that roughly 6--18\% of solar AR emergence participates in a non-random, hemispherically asymmetric nested component.

\subsection{Mixed-Parity Dynamo Interpretation}

Mixed-parity dynamo states provide a possible framework for interpreting hemispheric asymmetry in solar activity. Interaction between dipolar and quadrupolar magnetic-field components can enhance activity in one hemisphere while suppressing activity in the other \citep{Gallet2009, augustson:2015}. Temporal evolution of the large-scale low-order magnetic field over the solar cycle has also been examined observationally by \citet{DeRosa:2012}.

Figure~\ref{fig:quadupole} illustrates the temporal evolution of the signed axial harmonic coefficients during Solar Cycle 24 together with the activity-nest counts and hemispheric QBO residuals derived from detrended hemispheric sunspot-area variations. The HMI axial dipole ($\ell=1$, $m=0$), quadrupole ($\ell=2$, $m=0$), and octupole ($\ell=3$, $m=0$) coefficients are computed by projecting the HMI $B_r$ synoptic maps onto orthonormal spherical harmonics, following the same normalization convention used by \citet{DeRosa:2012}; this convention is applied uniformly across all degrees $\ell$, so the relative amplitudes of the dipole and quadrupole coefficients shown here are directly comparable. The WSO coefficients shown for comparison in Figure~\ref{fig:quadupole} instead use the Schmidt semi-normalization convention \citep{DeRosa:2012} and are rescaled for visual display only; they are not used to determine the shaded intervals or in any quantitative comparison in this section. The axial harmonic coefficients exhibit substantial temporal variability, including intervals during which the magnitude of the HMI axial quadrupole coefficient approaches or exceeds that of the HMI axial dipole coefficient. These intervals broadly overlap enhanced nesting activity, hemispheric QBO variability, and the polar-field reversal interval \citep{pishkalo2019}. The hemispheric QBO residuals and activity-nest counts exhibit moderate positive correlations in both hemispheres ($r = 0.54$ in the North and $r = 0.46$ in the South, see last two rows of Table~\ref{tab:nest_asymmetry}), with statistically significant $p$-values below $10^{-4}$. The temporal relationship between hemispheric QBO variability, nesting activity, and low-order magnetic-field evolution suggests that intermediate-timescale dynamo variability may contribute to the hemispheric organization of solar activity. A more detailed investigation of the relationship between QBO-like variability and activity nests is warranted in future work.

A mixed-parity magnetic field may be written schematically as
\begin{equation}
\vec{B} = D(t)\,\vec{B}_D + Q(t)\,\vec{B}_Q,
\end{equation}
where $\vec{B}_D$ and $\vec{B}_Q$ represent dipolar- and quadrupolar-parity components of the interior dynamo, and $D(t)$ and $Q(t)$ are their respective amplitudes, following the theoretical framework of \citet{Gallet2009} and \citet{augustson:2015}. In this framework, the relative amplitudes of the dipolar- and quadrupolar-parity dynamo families are thought to modulate hemispheric magnetic activity when the two components become comparable in strength. We note that the observed photospheric harmonic coefficients $B_1^0$ and $B_2^0$ are surface diagnostics of the emerged magnetic flux and do not directly constrain $D(t)$ and $Q(t)$ in the interior. The temporal coincidence between enhanced nesting activity, hemispheric asymmetry, and intervals of comparable $B_1^0$ and $B_2^0$ amplitude is therefore consistent with, though does not uniquely demonstrate, a mixed-parity interior dynamo state of the kind described by this framework. Coupling between dipolar- and quadrupolar-parity components can arise through equatorially asymmetric flows or source terms in the dynamo. Weak symmetry breaking in the velocity field has been shown to mix parity families and generate hemispherically localized magnetic fields through dipole--quadrupole interaction \citep{Gallet2009,augustson:2015}. In this context, cross-equatorial transport and hemispherically asymmetric flow variations \citep{sen2026a,sen2026b} may contribute to transient parity mixing that modulates the hemispheric organization of solar activity \citep{finley2024}. The concentration of statistically significant hemispheric asymmetry within activity nests may therefore provide a possible observational signature of temporally varying mixed-parity dynamo states.

\section{Conclusions}

We investigated the longitude--time distribution of solar AR emergence during Solar Cycles 22--24 and found statistically significant hemispheric asymmetry concentrated within the subset of ARs participating in activity nests. Monte Carlo realizations demonstrate that the observed nesting and hemispheric separation exceed expectations from random emergence alone. Wavelet analysis further shows that magnetic activity in the two hemispheres evolves in temporally offset bursts, contributing to the asymmetric organization of nested activity.

The fraction of ARs participating in nests is approximately 36--42\% across Cycles 22--24, although Monte Carlo realizations indicate that roughly half of this nesting fraction can arise by chance. The difference between the observed and Monte Carlo modeled nesting fraction therefore implies that approximately 18\% of AR emergence participates in non-random clustered activity. A more conservative lower bound based on the symmetry of the nest-member population suggests that approximately 6--8\% of all ARs participate in hemispherically asymmetric nesting behavior.

The disappearance of statistically significant asymmetry after removing nest-member ARs indicates that the background AR emergence process is broadly symmetric between hemispheres, while a smaller subset of ARs emerges in temporally and longitudinally localized hemispheric bursts. The concentration of asymmetry within activity nests may provide an observational signature of temporally varying mixed-parity dynamo states involving dipolar and quadrupolar magnetic-field families.

The temporal relationship between hemispheric QBO variability, nesting activity, and low-order magnetic-field evolution suggests that intermediate-timescale dynamo variability may contribute to hemispheric organization of solar activity.
A more detailed investigation of the relationship between QBO-like variability and activity nests is warranted in future work.
Future work combining activity nests, active longitudes, and large-scale magnetic-field evolution may help clarify the relationship between hemispheric asymmetry, QBO-like variability, and the global organization of solar magnetic activity.

\begin{acknowledgments}
This work was supported by the NASA DRIVE Center COFFIES grant 80NSSC20K0602. In addition, this project received support from the Whole Sun European Research Council Synergy grant under the European Union's Horizon 2020 research and innovation programme (grant agreement No. 810218).
\end{acknowledgments}

%\newpage
\bibliographystyle{aasjournal}
\bibliography{nests}

\end{document}